\documentclass[preprint,amsmath,amssymb,longbibliography]{revtex4-1}
\usepackage{txfonts}
\usepackage{amsfonts}
\usepackage{amsmath}
\usepackage{graphicx}
\usepackage{overpic}
\usepackage{color}
\usepackage{url}
\usepackage{hyperref}

\begin{document}

\title{The light pollution as a surrogate for urban population of the US cities}
\author{Felipe G. Operti$^{1}$, Erneson A. Oliveira$^{1,2}$, Humberto A.
Carmona$^1$, Javam C. Machado$^3$, Jos\'e S. Andrade Jr.$^{1}$}
%\email{erneson@fisica.ufc.br}
\affiliation{$^1$ Departamento de F\'isica, Universidade Federal do Cear\'a,
Fortaleza, Cear\'a, Brasil\\ $^2$ Programa de P\'os-Gradua\c{c}\~ao em
Inform\'atica Aplicada, Universidade de Fortaleza, Fortaleza, Cear\'a, Brasil\\
$^3$ Departamento de Computa\c{c}\~ao, Universidade Federal do Cear\'a,
Fortaleza, Cear\'a, Brasil}

\date{\today}

\begin{abstract}
{\bf We show that the definition of the city boundaries can have a dramatic
influence on the scaling behavior of the night-time light (NTL) as a function of
population (POP) in the US. Precisely, our results show that the arbitrary
geopolitical definition based on the Metropolitan/Consolidated Metropolitan
Statistical Areas (MSA/CMSA) leads to a sublinear power-law growth of NTL with
POP. On the other hand, when cities are defined according to a more natural
agglomeration criteria, namely, the City Clustering Algorithm (CCA), an
isometric relation emerges between NTL and population. This discrepancy is
compatible with results from previous works showing that the scaling behaviors
of various urban indicators with population can be substantially different for
distinct definitions of city boundaries. Moreover, considering the CCA
definition as more adequate than the MSA/CMSA one because the former does not
violate the expected extensivity between land population and area of their
generated clusters, we conclude that, without loss of generality, the CCA
measures of light pollution and population could be interchangeably utilized in
future studies.}
\end{abstract}
\maketitle

\noindent
{\large \bf Introduction}

More than 80\% of the world and more than 90\% of the US and European
populations live under light-polluted skies (exposition to light at night)
\cite{falchi2016}. Since the first electric-powered illumination in the second
half of the 19th century, the world has become covered by artificial electric
light, changing drastically the night view of the Earth from space. The
spreading of artificial electric light plays an important role on the duration
of the {\it productive day}, not only for working but also for recreational
activities. If in one hand the benefits of artificial light are quite evident,
on the other hand, scientific researches suggest that the exposition to light at
night could have adverse effects on both human and wildlife health
\cite{kerenyi1990, blask2005, reiter2006, navara2007, reiter2007, chepesiuk2009,
salgado-delgado2011, aube2013}. For example, in humans, the pineal and blood
melatonin rhythms are quickly disturbed by light pollution. Such studies argue
that the night light exposure have two major physiological effects: they disrupt
the circadian rhythms and suppress the production of melatonin
\cite{salgado-delgado2011}. This repeated suppression may have large
consequences for the mammals health. For instance, it was shown that the
suppression of the melatonin at night accelerates the metabolic activity and
growth of rat hepatoma \cite{navara2007} and human breast cancer
\cite{blask2005}. Moreover, the disruption of circadian rhythms made by the
exposure of light at night might plays a crucial role in the etiology of
depression \cite{salgado-delgado2011}.

The significant consequences of the exposure to night-time light (NTL) with the
fact that 54\% of world's population lives in urban areas stimulates the
interest in understanding how the light pollution evolves with the size of the
US cities \cite{un2014}. Bettencourt {\it et al.} found the cities in the US
exhibit three different types of allometric laws for urban indicators with
population size \cite{bettencourt2007}: (i) {\it Superlinear} . The superlinear
urban indicators increase proportionally more than the population of the cities.
Such behavior is intrinsically associated with the {\it social currency} of a
city, indicating that larger cities are associated with optimal levels of human
productivity and quality of life. Doubling the city size leads to a
larger-than-double increment in productivity and life standards
\cite{bettencourt2007, bettencourt2010a, Bettencourt2010b}. For example, wages,
income, growth domestic product (GDP), bank deposits, as well as rates of
invention measured by the number of patents and employment in creative sectors
show a superlinear behavior \cite{bettencourt2007}. (ii) {\it Linear} or {\it
isometric relation}. The increasing of the linear urban indicators is
proportional to the increasing of the population reflecting the common
individual human needs, like the number of jobs, houses, and water consumption
\cite{bettencourt2007}. (iii) {\it Sublinear}. The sublinear urban indicators
increase proportionally less than the population of the cities. This case is a
manifestation of the {\it economy of scale}. The sublinearity is found in the
number of gasoline stations, length of electrical cables, and road surfaces
(material and infrastructure) cases \cite{bettencourt2007}. From the results
shown by Bettencourt {\it et al.}, several studies have been carried out on the
allometry of urban indicators in different levels of human aggregation
\cite{melo2014, oliveira2014, bettencourt2016, caminha2017}. Following this aim,
we analyze and classify the allometric law between the NTL and the population of
the US cities.

Here, we use three geo-referenced dataset: the population dataset, the NTL
dataset and the Metropolitan/Consolidated Metropolitan Statistical Area
(MSA/CMSA). In order to define the boundaries of each US city, we use two
methods: the City Clustering Algorithm (CCA) \cite{makse1995, rozenfeld2008,
rozenfeld2011} and the MSA/CMSA \cite{censusbureau2014}. Finally, we find the
allometric scaling between the NTL and the population for the two applied
methods. Furthermore, to compare them, we analyze the allometric scaling between
area and population.

\noindent
{\large \bf Materials and Method}

\noindent
{\bf Population dataset (GPWv4)}

\begin{figure}[h!]
\centerline{\includegraphics[width=0.9\textwidth]{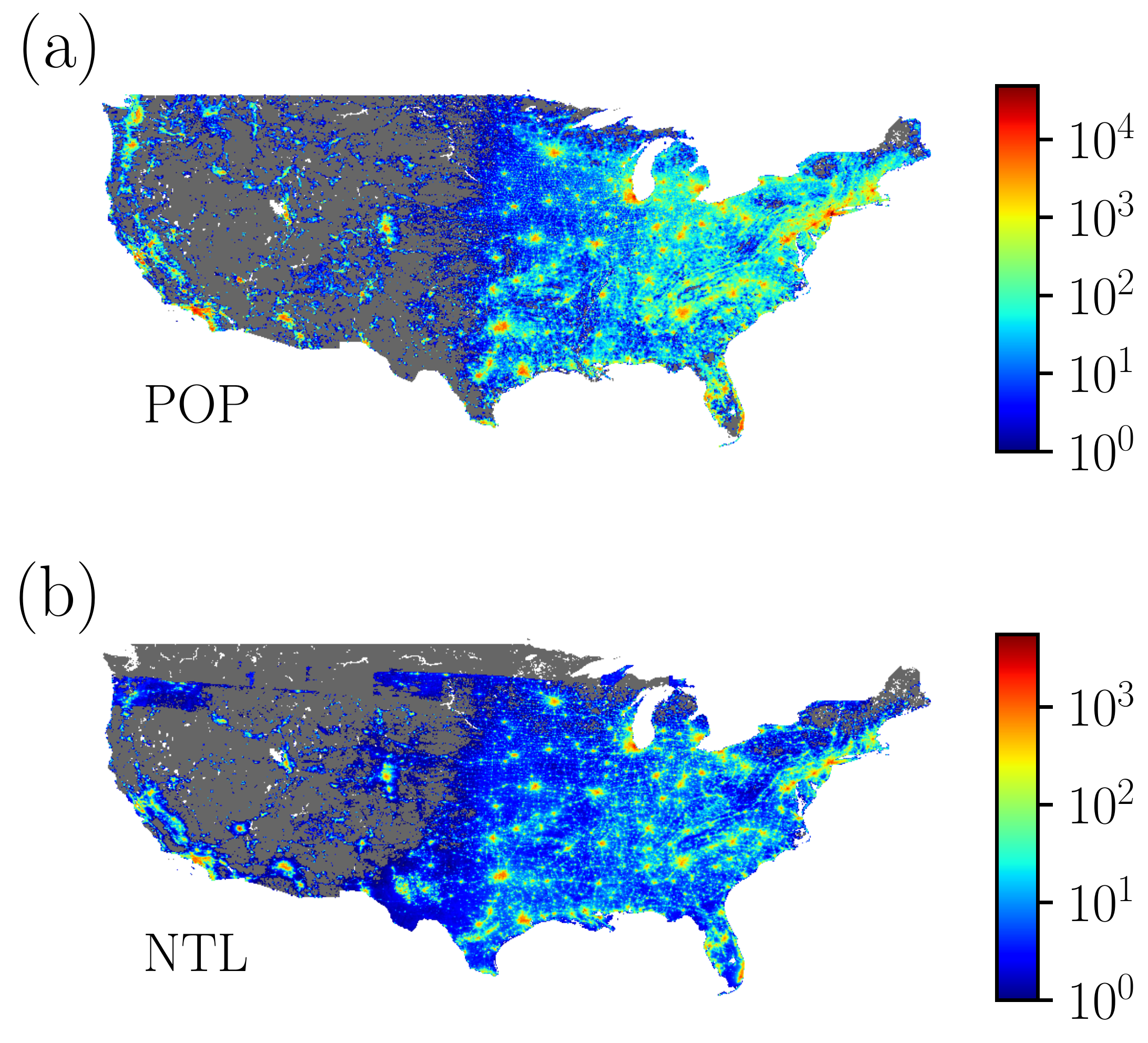}}
\caption{{\bf Datasets (on colors).} (a) The population dataset is defined as a
30 arc-second geolocated grid. It is obtained from the GPWv4 in logarithmic
scale for the year 2015 \protect\cite{doxsey-whitfield2015, sedac2016}. (b) The
NTL dataset is obtained through the night-time light radiance emission data from
the VIIRS DNB in $nW/cm^2/sr$ \protect\cite{mills2013, ncei2016, viirs2016}. It
is defined at the resolution of 15 arc-second grid in logarithmic scale for the
year 2015 (April).}
\label{fig1}
\end{figure}

The population dataset is extracted from the fourth version of the Gridded
Population of the World (GPWv4) \cite{doxsey-whitfield2015, sedac2016} from the
Center for International Earth Science Information Network (CIESIN) at the
Columbia University. The GPWv4 models the human population distribution on a
continuous surface at high resolution. Population input data is collected
through several censuses around the US, between 2005 and 2014. Data are provided
in grid form, where each cell is formed by 30 arc-second angles (approximately 1
km $\times$ 1 km at the Equator line). We use the US population count data,
measured in number of people, for the year 2015, as depicted in
Fig. \ref{fig1}a.

%%% FIG1 %%%
%%% FIG1 %%%

\noindent
The method successively introduced requires the population density of each grid
cell. Therefore, we calculated the area of ​​each grid cell dividing them into
two spherical triangles. The area of a spherical triangle with edges $a$, $b$
and $c$ is given by,
\begin{equation}\label{eq2}
A=R^2 E,
\end{equation}

\noindent
where $R=6,378.137$ km is the Earth's radius and the spherical excess $E$ is
defined by the following expression:
\begin{equation}\label{eq3}
E=4\tan^{-1}\left[\tan\left(\frac{s}{2}\right)\tan\left(\frac{s_a}{2}\right)\tan\left(\frac{s_b}{2}\right)\tan\left(\frac{s_c}{2}\right)\right]^{1/2}.
\end{equation}

\noindent
with $s=(a/R+b/R+c/R)/2$, $s_a=s-a/R$, $s_b=s-b/R$, and $s_c=s-c/R$. In this
context, the distance between two points, $i$ and $j$, on the Earth's surface is
calculated by,
\begin{equation}\label{eq4}
d_{ij}=R\theta,
\end{equation}

\noindent
with
\begin{equation}\label{eq5}
\theta=\cos^{-1}[\sin(y_i) \sin(y_j)+\cos(y_i)\cos(y_j)\cos(x_j-x_i)].
\end{equation}

\noindent
In this formalism, the values of $x_i$ ($x_j$) and $y_i$ ($y_j)$ are the
longitude and latitude, respectively, of the point $i$ ($j$), measured in
radians.

\noindent
{\bf Night-time light dataset (NTL)}

The NTL dataset is given by the night-time light radiance emission data from the
National Centers for Environmental Information (NCEI) \cite{ncei2016}. The NTL
dataset is defined by the monthly average of radiance, measured in $nW/cm^2/sr$,
using the night-time data from the scanning radiometer Visible Infrared Imaging
Radiometer Suite (VIIRS) Day/Night Band (DNB) \cite{mills2013, ncei2016,
viirs2016}. The VIIRS DNB data are processed and filtered in order to exclude
data impacted by the lunar illumination, lightning and cloud-cover, but they are
susceptible to other temporal lights, {\it e.g.} aurora, fires, and boats
\cite{mills2013, ncei2016}. Such data span through the entire globe with a
resolution of 15 arc-second (approximately 500 m $\times$ 500 m at the Equator
line) between the latitudes 75$^\circ$ North and 65$^\circ$ South. We use the US
data for the year 2015 (April), as shown in Fig. \ref{fig1}b.

\noindent
{\bf Metropolitan Statistical Area (MSA), Primary Metropolitan Statistical
Area (PMSA) and Consolidated Metropolitan Statistical Area (CMSA)}

The MSA are geographic entities with high degree of socioeconomic integration
and population over 50,000 people. The PMSA are quite similar to MSA, however
they present population over 1,000,000 people. The CMSA are metropolitan regions
defined by the agglomeration of some PMSA. They are all delineated by the Office
of Management and Budget (OMB) and provided by the US Census Bureau
\cite{censusbureau2014}.

\noindent
{\bf Data processing}

In order to superimpose the datasets, we perform two processes: (i) As the NTL
grid has a higher resolution than the GPWv4 grid, we sum the values of all NTL
grid cells, which their geolocated centers are within the same geolocated GPWv4
grid cell. Therefore, we produce a new NTL grid with the same positioning and
resolution of the GPWv4 dataset; (ii) For the MSA/CMSA case, we use the same
approach of (i), even though the MSA/CMSA are complex polygons. To deal with
this problem, we use the even-odd rule algorithm \cite{shimrat1962}. Thus, we
define the NTL value for each MSA/CMSA.

\noindent
{\bf City Clustering Algorithm (CCA)}

\begin{figure}[h!]
\centerline{\includegraphics[width=0.8\textwidth]{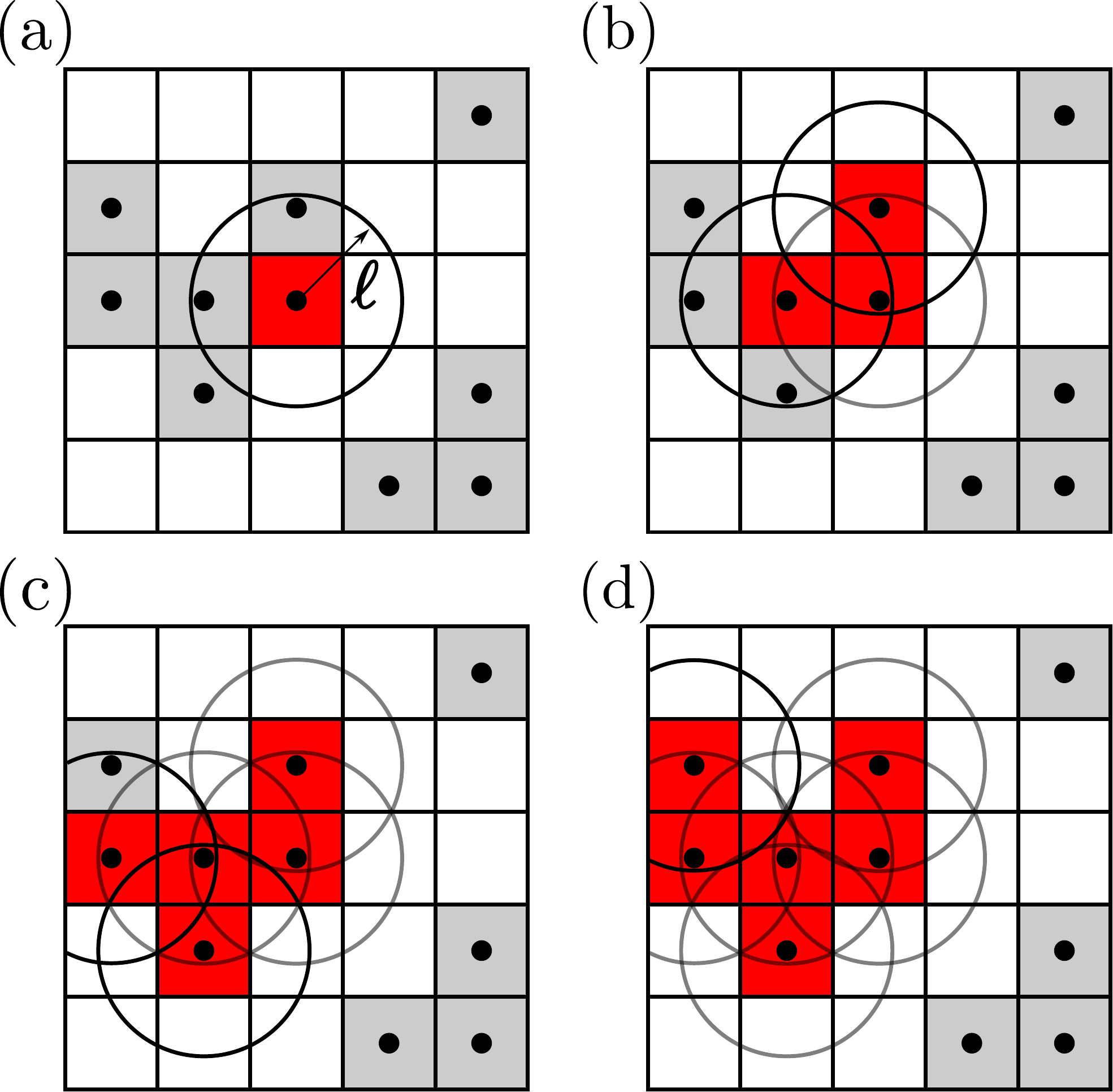}}
\caption{{\bf The CCA steps (on colors).} The grey and the the red cells are
populated ($D_i>D^*$). The small black circles are the geometric centers of each
populated cell. The red cells belong the same analyzed cluster. (a) First step:
the algorithm select a populated cell and draw a circle of radius $\ell$. (b)
Second step: the cells with the geometric centers inside the circles of radius
$\ell$ become a part of the red cluster and from their geometric center are
drawn others two circles of radius $\ell$. The circle of the first step is
showed in opaque black. (c) Third step: two more cells became part of the red
cluster and two more circles are drawn. (d) Fourth step: the last cell became
part of the red cluster. The entire cluster is determined and the algorithm will
start to analyze another cluster.}
\label{fig2}
\end{figure}

We define the boundaries of each US city by applying the CCA to the population
grid \cite{makse1995, rozenfeld2008, rozenfeld2011}. We use the continuum CCA
that depends on two parameters, namely, a population density threshold, $D^*$
and a cutoff length, $\ell$ \cite{rozenfeld2011}. For the $i$-th grid cell, the
population density $D_i$ is geo-referenced in its geometric center (shown as
small black circles in Fig. \ref{fig2}). If $D_i > D^*$, the $i$-th grid cell is
populated. In Fig. \ref{fig2} the populated cells are shown in grey and red.
Next, the algorithm selects a populated cell (red cell in Fig. \ref{fig2}a) and
aggregates in the same cluster all nearest populated cells which are within a
distance $\ell$ from each (red cells in Figs. \ref{fig2}b, \ref{fig2}c and
\ref{fig2}d). The Fig. \ref{fig2} shows the four steps to determine the red
cluster.

%%% FIG2 %%%
%%% FIG2 %%%

\noindent
{\large \bf Results}

We apply the CCA to the population grid varying $D^*$ (in $people/km^2$), from 0
to 10000, and $\ell$ (in $km$), from 1 to 20. For all pairs of parameters, we
find that it is possible to statistically correlate through power-law relations
the area and the population as well as the NTL and the population of the US
cities,

\begin{equation}\label{eq6}
\log(\text{AREA})=a+\alpha_{CCA} \log(\text{POP}),
\end{equation}
\begin{equation}\label{eq8}
\log (\text{NTL})=b+\beta_{CCA} \log (\text{POP}).
\end{equation}

\begin{figure}[h!]
\centerline{\includegraphics[width=1.0\textwidth]{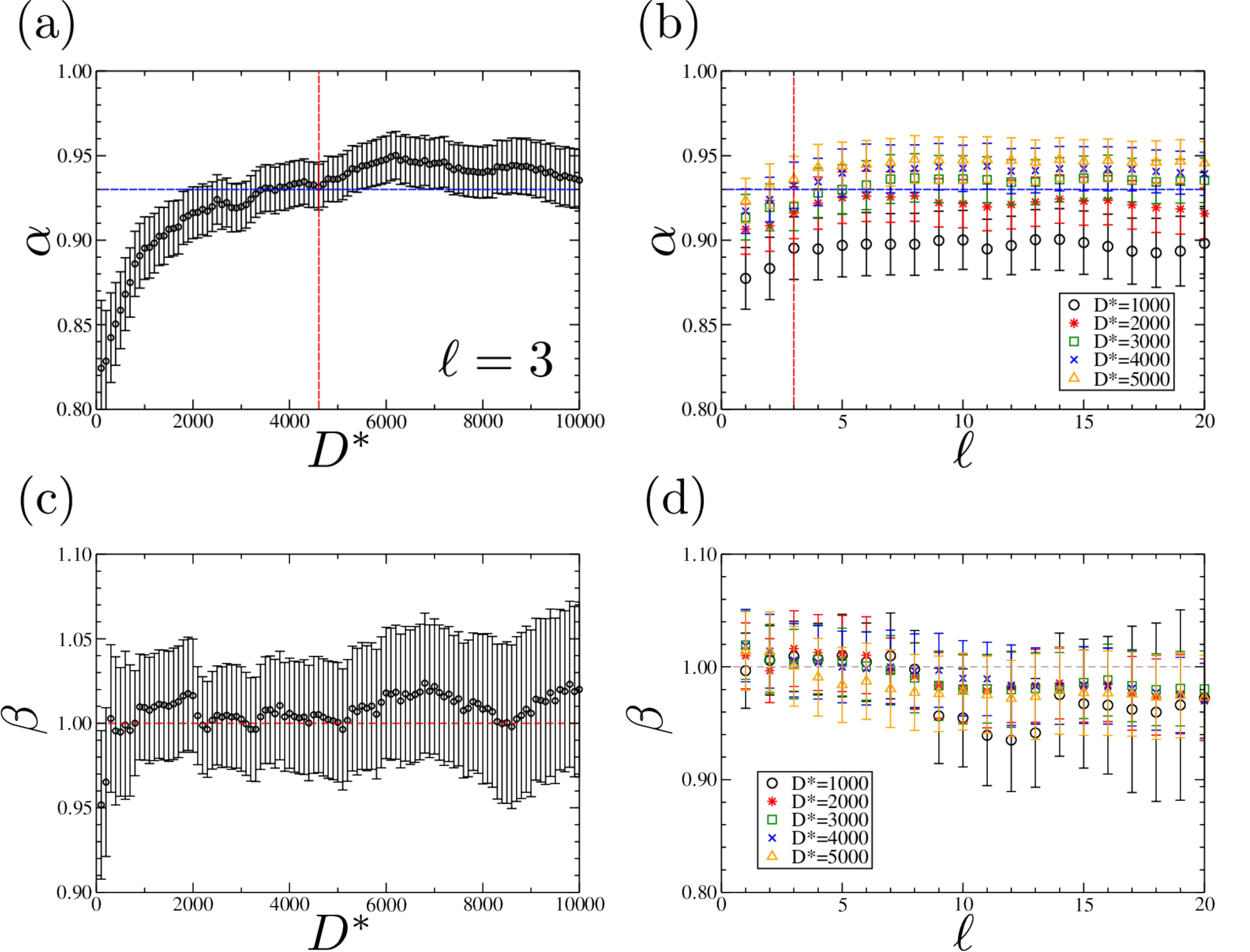}}
\caption{{\bf Allometric exponent $\alpha_{CCA}$ and $\beta_{CCA}$ as a function
of the parameter $D^*$ and $\ell$ (on colors).} (a) The exponent $\alpha_{CCA}$
as a function of $D^*$ for $\ell=3$ $km$. The parameter $D^*$ varies from 0 to
10000 $people/km^2$. For $D^* > 4000$ and $\ell=3$ the allometric exponent
$\alpha_{CCA}$ is between $0.93$ (dashed blue line) and $0.95$. For $D^*=4560$
$people/km^2$ (dashed red line) we observe the arising of five large cities in
US Northeast Coast. (b) The exponent $\alpha_{CCA}$ as a function of the CCA
parameter $\ell$ for $D^*=$1000, 2000, 3000, 4000, and 5000 $people/km^2$. We
find a plateau region after $\ell=3$ $km$, where $\alpha_{CCA}\approx 0.93$
(dashed blue line). (c) The figure shows the allometric exponent $\beta_{CCA}$
as a function of $D^*$ for $\ell=3$ $km$. The parameter $D^*$ varies from 0 to
10000 $people/km^2$. The dashed red line corresponds to $\beta=1$. (d) The
figure shows the allometric exponent $\beta_{CCA}$ as a function of the CCA
parameter $\ell$ for $D^*=$1000, 2000, 3000, 4000, and 5000 $people/km^2$. The
dashed brown line corresponds to $\beta=1$.}
\label{fig3}
\end{figure}

The exponents $\alpha_{CCA}$ and $\beta_{CCA}$ are obtained through Ordinary
Least Square (OLS) \cite{montgomery2015} fitting to the data for different
values of the parameters $D^*$ and $\ell$. The ranges of compatibility and the
consistency of the CCA technique are investigated in Figs. \ref{fig3}a-d.

%%% FIG 3 %%%
%%% FIG 3 %%%

\begin{figure}[h!]
\centerline{\includegraphics[width=1.0\textwidth]{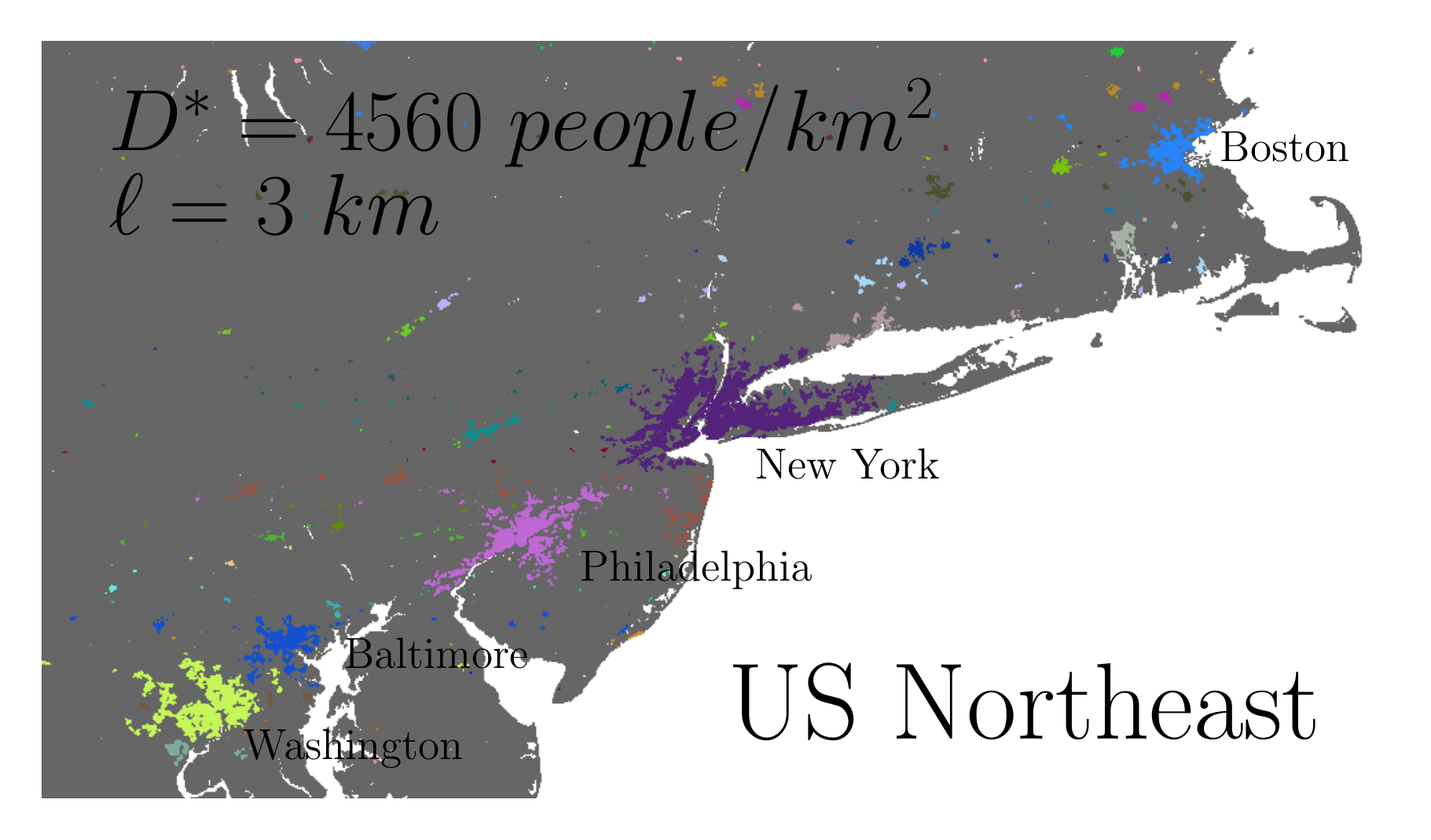}}
\caption{{\bf Application of CCA to the US Northeast region (on colors).} We use
the CCA parameters $D^*=$4560 $people/km^2$ and $\ell=3$ $km$. The clusters of
different colors identify different urban agglomerations. Essentially, we
distinguish five famous cities such as Boston (light blue), New York (purple),
Philadelphia (pink), Baltimore (blue), and Washington D.C. (light green).}
\label{fig4}
\end{figure}

Indeed, the definition of the parameters $D^*$ and $\ell$ of the CCA affects the
dimension and the geometry of the cities, but from the Figs. \ref{fig3}c and
\ref{fig3}d, it can be seen that it does not affect the allometric exponent
$\beta_{CCA}$. Here, our starting strategy is to determine a range of parameters
$D^*$ and $\ell$ for which the relation between area and population is isometric
\cite{rozenfeld2008, rozenfeld2011, oliveira2014, caminha2017}. We find that for
$D^* > 4000$ and $\ell=3$ the allometric exponent $\alpha_{CCA}$ is between
$0.93$ and $0.95$ and we consider this relation approximately linear. Inside
this range, we analyze the result of the CCA using $D^*=4560$ and $\ell=3$,
where the five larger cities in the US Northeast Coast naturally emerge, as
depicted in Fig. \ref{fig4}. We believe that, the lack of an exactly linearity,
also inside this range, is due to the high density of some downtowns,
specifically, of the most populated urban centers of the US Northeast Coast.

%%% FIG4 %%%
%%% FIG4 %%%

\begin{figure}[h!]
\centerline{\includegraphics[width=1\textwidth]{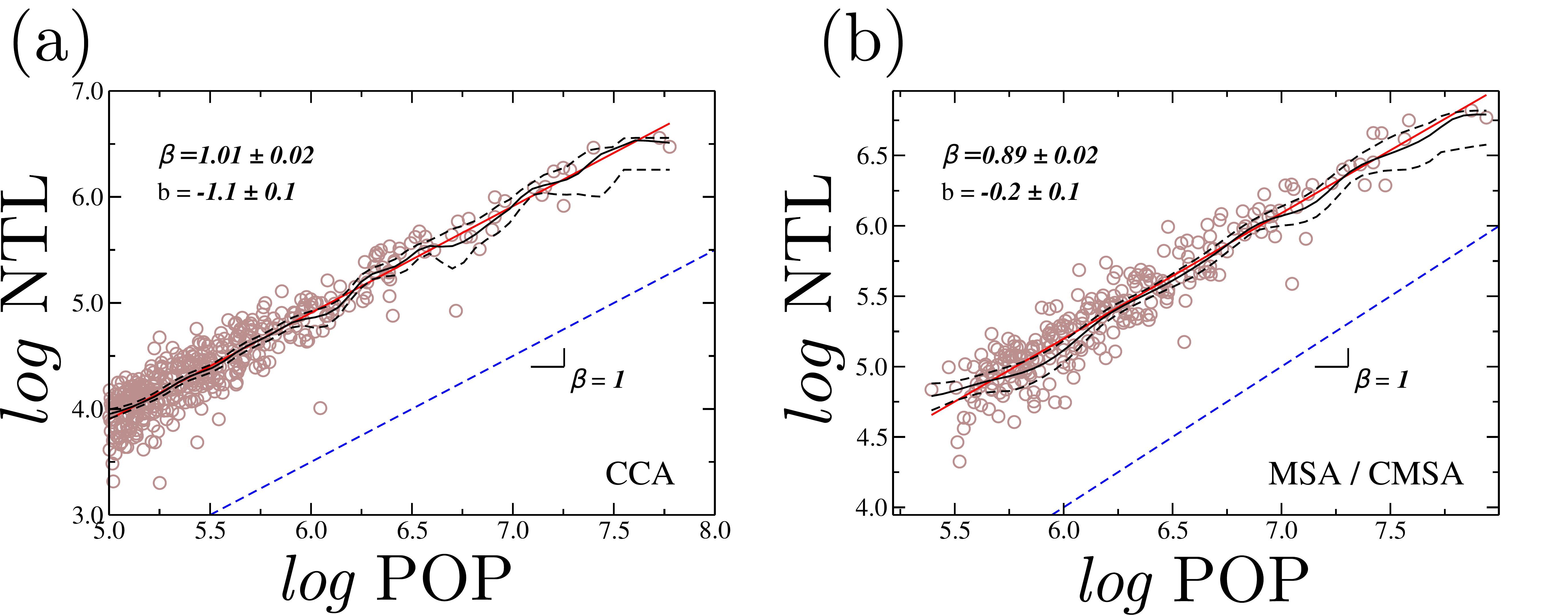}}
\caption{{\bf Allometric exponent $\alpha$ applying the CCA and using the
MSA/CMSA definitions (on colors).} (a) The figure shows the allometric scaling
law in Eq. \ref{eq6} and its allometric scaling exponent $\alpha_{CCA}=0.93 \pm
0.01$ using CCA parameters $D^{\ast}=$4560 $people/km^2$ and $l=3$ $km$. The red
line is the OLS result, and the solid black line is the N-W estimator. The
dashed black lines show the 95\% confidence bands of the N-W. The dashed blue
line corresponds to $\alpha=1$. (b) The figure shows the allometric scaling
exponent $\alpha_{MSA/CMSA}=0.49 \pm 0.03$ using the MSA/CMSA definitions. The
red line is the OLS result, and the solid black line is the N-W estimator. The
dashed black lines show the 95\% confidence bands of the N-W. The dashed blue
line corresponds to $\alpha=1$.}
\label{fig5}
\end{figure}

\begin{figure}[h!]
\centerline{\includegraphics[width=1\textwidth]{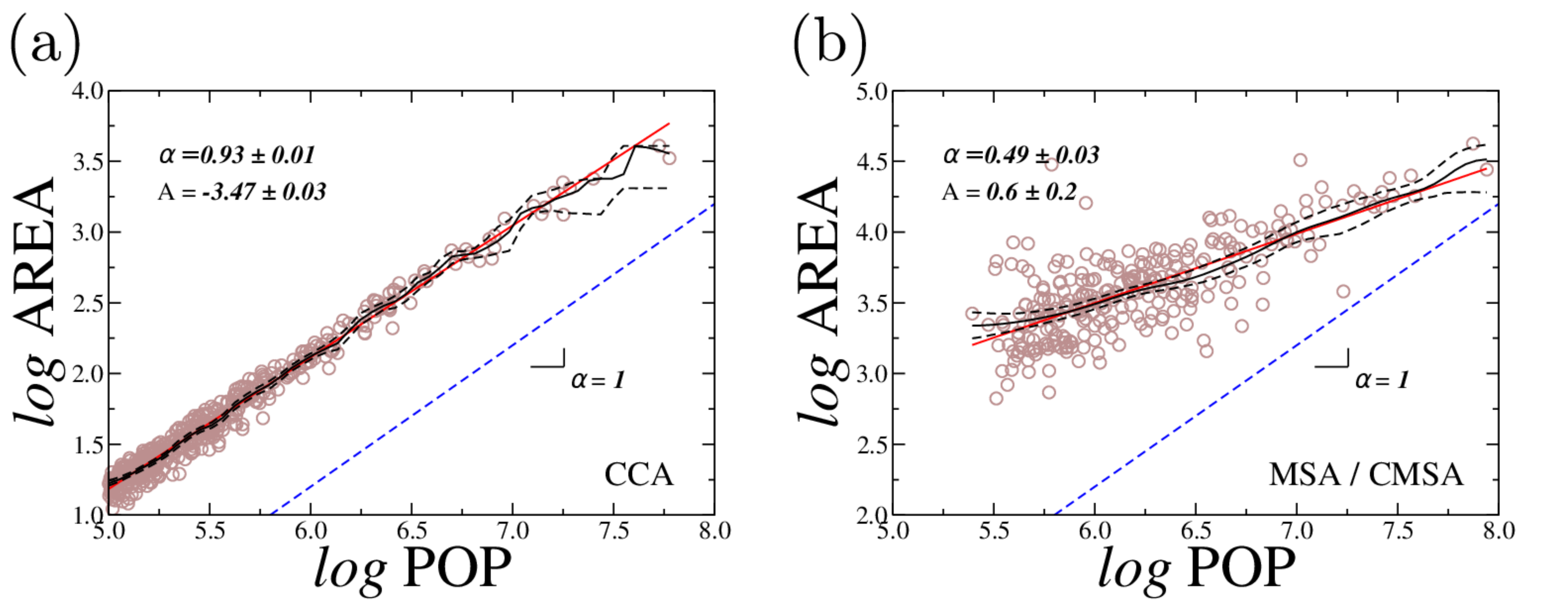}}
\caption{{\bf NTL versus population using the CCA and the MSA/CMSA definitions
(on colors).} (a) NTL versus population using CCA parameters $D^{\ast}=$4560
$people/km^2$ and $l=3$ $km$. The graph shows a linear relation between the NTL
measured in $nW/cm^2/sr$ and the population with allometric scaling exponent
$\beta_{CCA} =1.01\pm0.02$ ($R^2=0.88$). The solid red line is the linear
regression obtained using the OLS method. The solid black line is the N-W
estimator and the dashed black lines show the lower and the upper confidence
interval (95\%)\protect\cite{nadaraya1964, watson1964}. The dashed blue line
corresponds to $\beta=1$. (b) NTL versus population using MSA/CMSA. The graph
shows a sublinear relation between the NTL, measured in $nW/cm^2/sr$, and the
population with allometric scaling exponent $\beta =0.89\pm0.02$ ($R^2=0.89$).
The red line is the linear regression and the black line is the N-W estimator.
The dashed black lines show the 95\% confidence band of the N-W.}
\label{fig6}
\end{figure}

For the pair of parameters, $D^*=4560$ and $\ell=3$, we find a allometric
exponent $\alpha_{CCA}=0.93 \pm 0.01$ (Figs. \ref{fig5}a) and a linear scaling
between NTL and the population with exponent $\beta_{CCA}=1.01\pm 0.02$ (Figure
\ref{fig6}a). Alternatively, others parameters inside this range could be
analyzed without affecting the allometric exponent $\beta_{CCA}$ (as shown in
Figs. \ref{fig3}c and \ref{fig3}d).

%%% FIG5 %%%
%%% FIG5 %%%

By analyzing the allometric scaling of the NTL with the population of the US
cities using the MSA/CMSA (Fig. \ref{fig6}b), we obtain the allometric exponent
$\beta_{MSA/CMSA} =0.89\pm 0.02$. Such an exponent characterizes a sublinear
relation between the NTL and the population, in contrast with the CCA result.

As shown in Fig. \ref{fig5}b, the sublinear scaling behavior of the MSA/CMSA
areas as a function of their corresponding populations,
$\alpha_{MSA/CMSA}=0.49\pm0.03$, clearly suggests that this might not be the
most adequate definition of a city agglomerate to be adopted in our study.

%%% FIG6 %%%
%%% FIG6 %%%

\begin{figure}[h!]
\centerline{\includegraphics[width=1.0\textwidth]{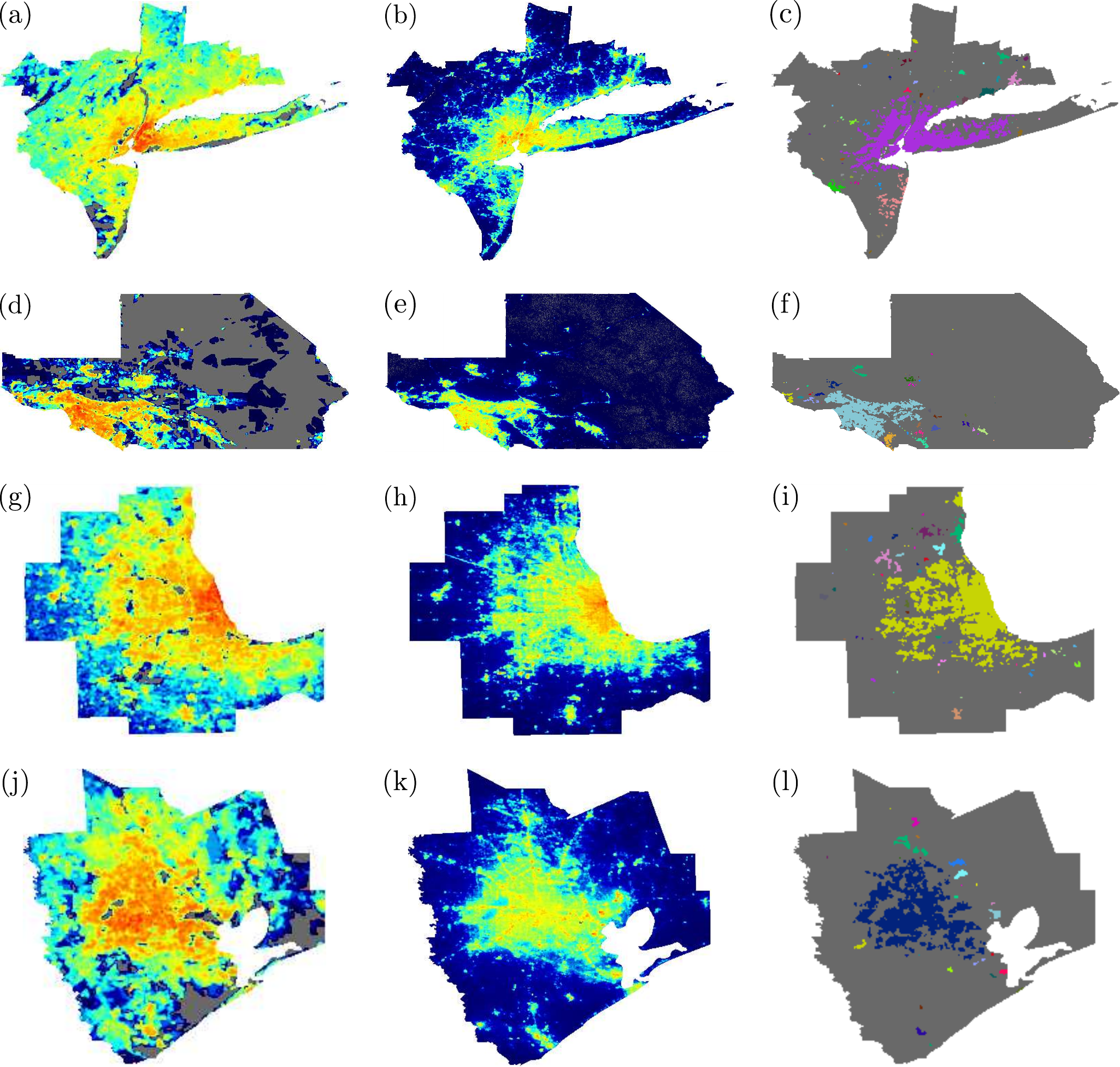}}
\caption{{\bf Comparison between the CCA and MSA/CMSA (on colors).} Figures (a),
(d), (g) and (j) are the human population grid in logarithmic scale obtained
from the GPWv4 for the year 2015\protect\cite{sedac2016,doxsey-whitfield2015}.
Figures (b), (e), (h) and (k) are the NTL measured in logarithmic scale with
units $nW/cm^2/sr$ obtained through the night-time light radiance emission data
from the VIIRS DNB\protect\cite{mills2013, ncei2016, viirs2016}. In figures (c),
(f), (i) and (l) we show the CCA clusters obtained using the CCA parameters
$D^{\ast}=$4560 $people/km^2$ and $l=3$ $km$ of the CMSA of: New York-Northern
New Jersey-Long Island (NY, NJ, CT, PA), Los Angeles-Riverside-Orange County
(CA), Chicago-Gary-Kenosha (IL,IN,WI) and Houston-Galveston-Brazoria (TX). The
figures show the discrepancy between the area estimated by the MSA/CMSA and the
area delimited by the CCA.}
\label{fig7}
\end{figure}

As indicated by Oliveira {\it et al.} \cite{oliveira2014}, the arbitrary
geopolitical concept behind the MSA/CMSA seems to overestimate the natural
limits of urban areas. In order to illustrate this fact, we show in Fig.
\ref{fig7} the MSA/CMSA of the five most populated US regions, namely, New
York-Northern New Jersey-Long Island (NY, NJ, CT, PA), Los
Angeles-Riverside-Orange County (CA), Chicago-Gary-Kenosha (IL,IN,WI) and
Houston-Galveston-Brazoria (TX). The first and second columns show respectively
the detailed maps of the population and the NTL datasets. The third column
exhibits the cities defined by the CCA with $D^*$=4560 and $\ell$=3, as well as
the discrepancy between the urban areas belonging to MSA/CMSA and CCA.

%%% FIG7 %%%
%%%% FIG7 %%%

\noindent
{\large \bf Conclusions}

We analyzed the allometric scaling behavior of the NTL as a function of the
population of the US cities. Our results corroborate previous works showing that
the scaling behaviors of urban indicators with population can be substantially
different for distinct definitions of city boundaries. Precisely, using the
MSA/CMSA definition, we found a sublinear allometric scaling exponent
$\beta_{MSA/CMSA}=0.89\pm002$. Applying the CCA, we found an exponent
$\beta_{CCA}=1.01\pm 0.02$ which indicates an isometric relation between the
light pollution and the population of the US urban agglomerations, in clear
contrast with the exponent obtained using the MSA/CMSA. Considering the
consistency of the CCA definition in terms of the extensivity between land
population and area of their generated clusters, as demonstrated in previous
studies for other urban indicators \cite{oliveira2014}, we come to the
conclusion that the proportionality between light pollution and population is
indeed correct, as intuitively expected \cite{li2015}. Under this framework and
without loss of generality, it is therefore plausible to utilize NTL as a
surrogate for city population in future studies.

The isometric relation between NTL and population of the US urban agglomeration,
obtained applying the CCA, imply that small and large cities are proportionally
indistinguishable in terms of light pollution. In other words, there is no {\it
economy of scale} or sublinearity concerning the NTL in US cities. Our result
shows that a growth of the US cities will aggravate the light pollution and
therefore the possible negative effects of the light pollution for the humans
and the wildlife health.

\noindent
{\large \bf Acknowledgments}

We gratefully acknowledge CNPq, CAPES, FUNCAP and the National Institute of
Science and Technology for Complex Systems in Brazil for financial support. We
especially thank our colleagues and friends H. P. M. Melo and T. A. Amor for the
help and the discussions.


\noindent
{\large \bf References}

\begin{thebibliography}{10}
\bibitem{falchi2016} F.~Falchi, P.~Cinzano, D.~Duriscoe, C.~C.~M. Kyba, C.~D.
Elvidge, K.~Baugh, B.~A. Portnov, N.~A. Rybnikova, R.~Furgoni, {The new world
atlas of artificial night sky brightness}, Science Advances 2~(6) (2016)
e1600377.

\bibitem{kerenyi1990}
N.~A. Kerenyi, E.~Pandula, G.~Feuer, {Why the incidence of cancer is increasing:
The role of light pollution}, Medical Hypotheses 33~(2) (1990) 75.

\bibitem{blask2005}
D.~E. Blask, G.~C. Brainard, R.~T. Dauchy, J.~P. Hanifin, L.~K. Davidson, J.~A.
Krause, L.~A. Sauer, M.~A. Rivera-Bermudez, M.~L. Dubocovich, S.~A. Jasser,
D.~T. Lynch, M.~D. Rollag, F.~Zalatan, {Melatonin-depleted blood from
premenopausal women exposed to light at night stimulates growth of human breast
cancer xenografts in nude rats}, Cancer Research 65~(23) (2005) 11174.

\bibitem{reiter2006}
R.~J. Reiter, F.~Gultekin, L.~C. Manchester, D.~Tan, {Light pollution, melatonin
suppression and cancer growth}, Journal of Pineal Research 40~(4) (2006) 357.

\bibitem{navara2007}
K.~J. Navara, R.~J. Nelson, {The dark side of light at night: Physiological,
epidemiological, and ecological consequences}, Journal of Pineal Research 43~(3)
(2007) 215.

\bibitem{reiter2007}
R.~J. Reiter, D.~Tan, A.~Korkmaz, T.~C. Erren, C.~Piekarski, H.~Tamura, L.~C.
Manchester, {Light at Night, Chronodisruption, Melatonin Suppression, and Cancer
Risk: A Review}, Critical Reviews in Oncogenesis 13~(4) (2007) 303.

\bibitem{chepesiuk2009}
R.~Chepesiuk, {Missing the Dark: Health Effects of Light Pollution},
Environmental Health Perspectives 117~(1) (2009) A20.

\bibitem{salgado-delgado2011}
R.~Salgado-Delgado, A.~{Tapia Osorio}, N.~Saderi, C.~Escobar, {Disruption of
circadian rhythms: A crucial factor in the etiology of depression}, Depression
Research and Treatment 2011~(839743) (2011) 1.

\bibitem{aube2013}
M.~Aub{\'e}, J.~Roby, M.~Kocifaj, {Evaluating Potential Spectral Impacts of
Various Artificial Lights on Melatonin Suppression, Photosynthesis, and Star
Visibility}, PLoS ONE 8~(7) (2013) 1.

\bibitem{un2014}
{United Nations},
\href{http://www.un.org/en/development/desa/news/population/world-urbanization-prospects-2014.html}{{World's
population increasingly urban with more than half living in urban areas}},
accessed: 2017-06-01 (2014). \url{http://www.un.org/en/development/desa/news/population/world-urbanization-prospects-2014.html}

\bibitem{bettencourt2007}
L.~M.~A. Bettencourt, J.~Lobo, D.~Helbing, C.~K{\"u}hnert, G.~B. West, {Growth,
innovation, scaling, and the pace of life in cities}, Proceedings of the
National Academy of Sciences of the United States of America 104~(17) (2007)
7301.

\bibitem{bettencourt2010a}
L.~M.~A. Bettencourt, G.~B. West, {A unified theory of urban living}, Nature
467~(7318) (2010) 912.

\bibitem{Bettencourt2010b}
L.~M.~A. Bettencourt, J.~Lobo, D.~Strumsky, G.~B. West, {Urban scaling and its
deviations: Revealing the structure of wealth, innovation and crime across
cities}, PLoS ONE 5~(11) (2010) 20.

\bibitem{melo2014}
H.~P.~M. Melo, A.~A. Moreira, {\'E}.~Batista, H.~A. Makse, J.~S. Andrade,
{Statistical signs of social influence on suicides}, Scientific reports 4~(6239)
(2014) 1.

\bibitem{oliveira2014}
E.~A. Oliveira, J.~S. Andrade, H.~A. Makse, {Large cities are less green},
Scientific reports 4~(4235) (2014) 1.

\bibitem{bettencourt2016}
L.~M.~A. Bettencourt, J.~Lobo, {Urban scaling in Europe}, Journal of The Royal
Society Interface 13~(116) (2016) 1.

\bibitem{caminha2017}
C.~Caminha, V.~Furtado, T.~H.~C. Pequeno, C.~Ponte, H.~P.~M. Melo, E.~A.
Oliveira, J.~S. Andrade, {Human mobility in large cities as a proxy for crime},
PLoS ONE 12~(2) (2017) e0171609.

\bibitem{makse1995}
H.~A. Makse, S.~Havlin, H.~E. Stanley, {Modelling urban growth patterns}, Nature
377~(6550) (1995) 608.

\bibitem{rozenfeld2008}
H.~D. Rozenfeld, D.~Rybski, J.~S. Andrade, M.~Batty, H.~E. Stanley, H.~A. Makse,
{Laws of population growth.}, Proceedings of the National Academy of Sciences of
the United States of America 105~(48) (2008) 18702.

\bibitem{rozenfeld2011}
H.~D. Rozenfeld, D.~Rybski, X.~Gabaix, H.~A. Makse, {The area and population of
cities: New insights from a different perspective on cities}, American Economic
Review 101~(5) (2011) 2205.

\bibitem{censusbureau2014}
{US Census Bureau}, \href{http://www.census.gov}{{Cartographic Boundary Files}},
accessed: 2017-06-01 (2014). \url{http://www.census.gov}

\bibitem{doxsey-whitfield2015}
E.~Doxsey-Whitfield, K.~MacManus, S.~B. Adamo, L.~Pistolesi, J.~Squires,
O.~Borkovska, S.~R. Baptista, {Taking Advantage of the Improved Availability of
Census Data: A First Look at the Gridded Population of the World, Version 4},
Papers in Applied Geography 1~(3) (2015) 226.

\bibitem{sedac2016}
{Socioeconomic data and application center (SEDAC)},
\href{http://sedac.ciesin.columbia.edu}{{Gridded Population of the World,
Version 4 (GPWv4)}}, accessed: 2017-06-01 (2016).
\url{http://sedac.ciesin.columbia.edu}

\bibitem{mills2013}
S.~Mills, S.~Weiss, C.~Liang, {VIIRS day/night band (DNB) stray light
characterization and correction}, SPIE Optical Engineering Applications
8866~(88661P) (2013) 1.

\bibitem{ncei2016}
{National Centers for Environmental Information (NCEI)},
\href{http://www.ngdc.noaa.gov/eog/viirs.html}{{Visible Infrared Imaging
Radiometer Suite (VIIRS)}}, accessed: 2017-06-01 (2016).
\url{http://www.ngdc.noaa.gov/eog/viirs.html}

\bibitem{viirs2016}
{National Aeronautics and Space Administration (NASA)},
\href{http://npp.gsfc.nasa.gov/viirs.html}{{Visible Infrared Imaging Radiometer
Suite (VIIRS)}}, accessed: 2017-06-01 (2016).
\url{http://npp.gsfc.nasa.gov/viirs.html}

\bibitem{shimrat1962}
M.~Shimrat, {Algorithm 112: Position of point relative to polygon},
Communications of the ACM 5~(8) (1962) 434.

\bibitem{montgomery2015}
D.~C. Montgomery, E.~A. Peck, G.~G. Vining, {Introduction to linear regression
analysis}, John Wiley \& Sons, 2015.

\bibitem{nadaraya1964}
E.~A. Nadaraya, {On estimating regression}, Theory of Probability \& Its
Applications 9~(1) (1964) 141.

\bibitem{watson1964}
G.~S. Watson, {Smooth regression analysis}, Sankhy{\=a}: The Indian Journal of
Statistics, Series A (1964) 359.

\bibitem{li2015}
X.~Li, X.~Wang, J.~Zhang, L.~Wu, {Allometric scaling, size distribution and
pattern formation of natural cities}, Palgrave Communications 1~(15017) (2015)
1.

\end{thebibliography}
\end{document}